\documentclass[a4paper,11pt]{article}
\usepackage{pos}
\usepackage{siunitx}
\DeclareSIUnit{\PE}{PE}
\DeclareSIUnit{\SPS}{SPS}
\usepackage{placeins}

\title{Performance of the D-Egg Optical Sensor\\for the IceCube Upgrade}
 \ShortTitle{D-Egg Performance}

\author{The IceCube Collaboration
\\{\normalsize \normalfont(a complete list of authors can be found at the end of the proceedings)}\vspace{1ex}}

\emailAdd{colton.hill@icecube.wisc.edu}
\emailAdd{mmeier@icecube.wisc.edu}
\emailAdd{ryo.nagai@icecube.wisc.edu}
\emailAdd{kenichikin@icecube.wisc.edu}
\emailAdd{shimizu@hepburn.s.chiba-u.ac.jp}
\emailAdd{aya@icecube.wisc.edu}
\emailAdd{syoshida@icecube.wisc.edu}
\emailAdd{tba109@psu.edu}
\emailAdd{jim.braun@icecube.wisc.edu}
\emailAdd{atfienberg@psu.edu}
\emailAdd{jeff.weber@icecube.wisc.edu}

\abstract{New optical sensors called the ``D-Egg'' have been developed for cost-effective instrumentation for the IceCube Upgrade. With two 8-inch high quantum efficient photomultiplier tubes (PMTs), they offer increased effective photocathode area while retaining as much of the successful IceCube Digital Optical Module design as possible. Mass production of D-Eggs has started in 2020. By the end of 2021,  there will be 310 D-Eggs produced with 288 deployed in the IceCube Upgrade. The D-Egg readout system uses advanced technologies in electronics and computing power. Each of the two PMT signals is digitised using ultra-low-power 14-bit ADCs with a sampling frequency of \SI{240}{\mega\SPS}, enabling seamless and lossless event recording from single-photon signals to signals exceeding \SI{200}{\PE} within \SI{10}{\nano\second}, as well as flexible event triggering. In this paper, we report the single photon detection performance as well as the multiple photon recording capability of D-Eggs from the mass production line which have been evaluated with the built-in data acquisition system.

\vspace{4mm}
{\bfseries Corresponding authors:}
Colton~Hill$^{1*}$, Maximillian~Meier$^{1}$, Ryo~Nagai$^{1}$, Ken'ichi~Kin$^{1}$, Nobuhiro~Shimizu$^{1}$, Aya~Ishihara$^{1}$, Shigeru~Yoshida$^{1}$, Tyler~Anderson$^{2}$, Jim~Braun$^{3}$, Aaron~Fienberg$^{2}$, Jeff~Weber$^{3}$ \\
{\itshape 1~Dept. of Physics and Institute for Global Prominent Research, Chiba University, Japan}\\
{\itshape 2~Dept. of Astronomy and Astrophysics, Pennsylvania State University, University Park, USA}\\
{\itshape 3~Dept. of Physics and Wisconsin IceCube Particle Astrophysics Center, University of Wisconsin{\textendash}Madison, USA}\\[2mm]
$^*$ Presenter
}

\FullConference{37$^{\textrm{th}}$ International Cosmic Ray Conference (ICRC 2021)\\
		July 12th -- 23rd, 2021\\
		Online -- Berlin, Germany}

\begin{document}
\maketitle


\section{Introduction}
The IceCube Neutrino Observatory, located at the geographic South Pole, is composed of strings of optical sensors inserted into the Antarctic ice-sheet. Neutrino interactions in the ice create charged particles, which traverse the ice at high energies emitting Cherenkov light, whose emission spectrum peaks in the UV region. The in-ice digital optical modules (DOMs) detect this Cherenkov light using a single 10-inch photomultiplier tube (PMT). To enhance future physics results from IceCube, a number of new optical modules will be deployed at the South Pole as a part of the coming IceCube Upgrade project. 

One of these new optical modules, D-Egg, has been developed as the primary detector for the IceCube Upgrade, set to be constructed in the 2022/2023 South Pole season~\cite{IcUpgrade}. 
A total of \num{310}~\mbox{D-Eggs} will have been assembled by the end of 2021, of which \num{288} are planned for deployment at the South Pole. Innovations in the module design, in addition to optimisation of the glass housing size to reduce deployment costs (drilling), has made the D-Eggs both high-efficiency and cost-effective choices for the IceCube Upgrade.

\begin{figure}[htb]
\centering
\includegraphics[height=0.3\textheight,trim={{1.3\textwidth} 0 {1.3\textwidth} 0},clip]{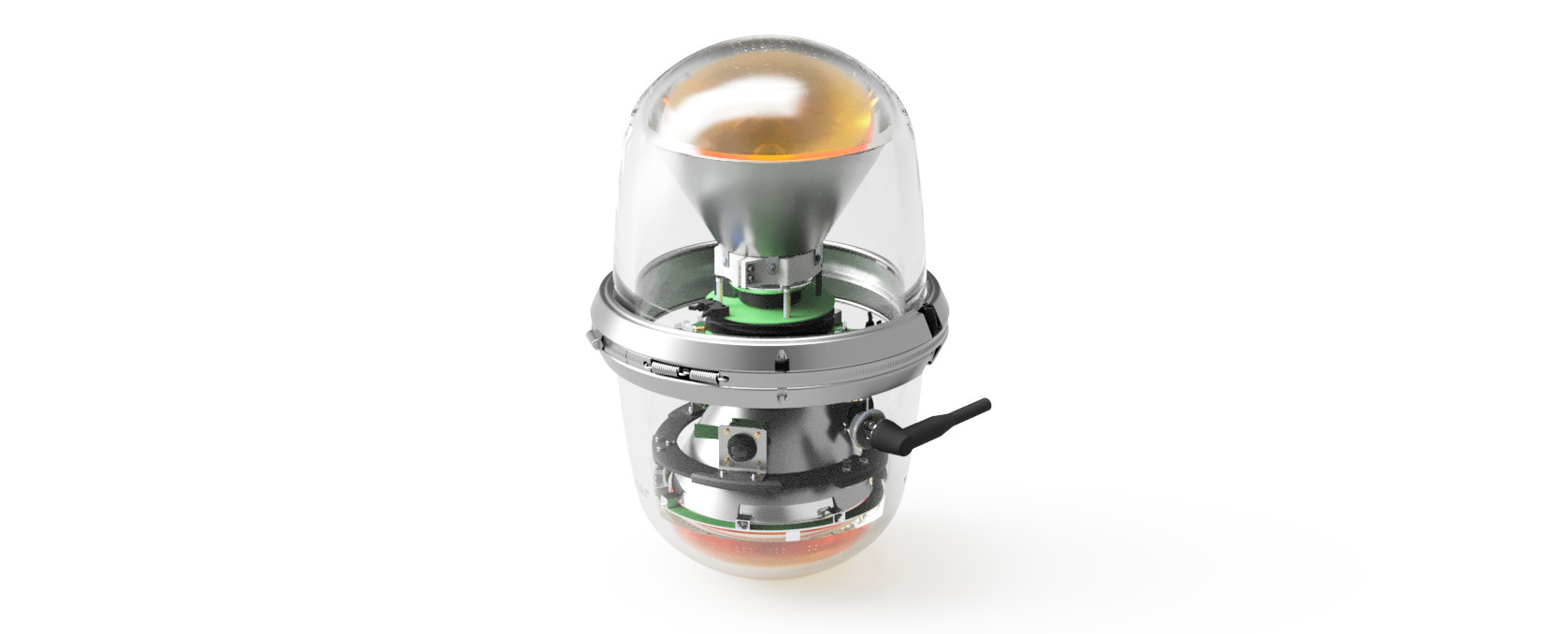}~
\includegraphics[height=0.3\textheight,trim={{1.3\textwidth} 0 {1.3\textwidth} 0},clip]{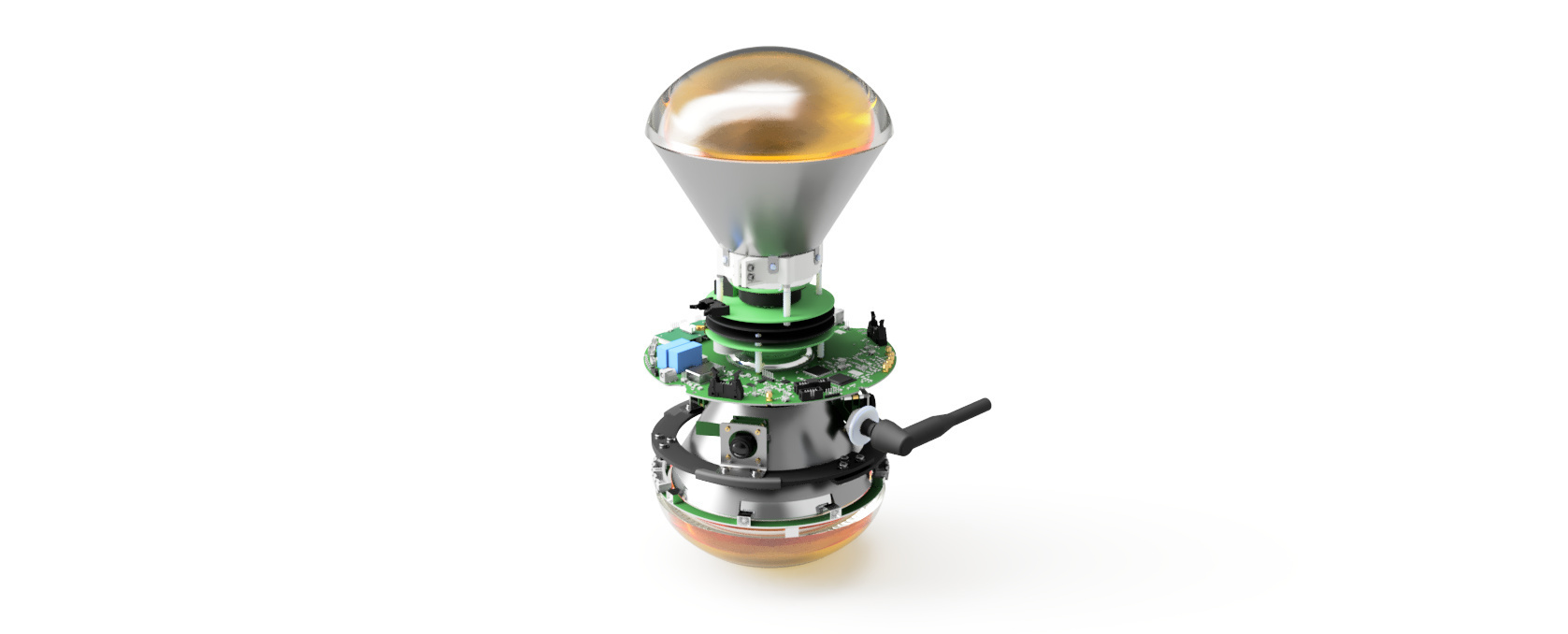}
\caption{Figure of the D-Egg module designed and built for the IceCube Upgrade. On the left is the D-Egg, sealed in the UV-transparent glass housing with the waistband harness around the equator, which will be used during deployment. Each PMT is carefully fixed in-place in either an upper or lower glass hemisphere by UV-transparent optical coupling gel. The figure on the right more cleanly shows the inside of the D-Egg, including the mainboard, calibration devices (cameras, LED flashers), and magnetic shielding.}
\label{tmp_degg}
\end{figure}

As opposed to the current generation, single-PMT IceCube DOMs, D-Eggs consist of \textit{two} high quantum efficient 8-inch PMTs. The two PMTs are accompanied by high voltage supply bases, a front-end circuit board (mainboard), magnetic shielding, and calibration devices. 
All components are housed inside an ellipsoidal pressure-resistant vessel made of UV-transparent borosilicate glass, which has increased UV transmissivity when compared to the IceCube DOM vessel. A figure of an assembled D-Egg can be seen in Fig.~\ref{tmp_degg}. 

IceCube is sensitive to neutrinos originating from both the Northern and Southern Skies, and by installing an upward and downward facing PMT (Hamamatsu Photonics R5912-100-70), the detection uniformity for each module improves significantly.
The mainboard is responsible for handling waveforms from the PMTs, controlling all electronic components inside the glass vessel,  
and communicating with the data acquisition (DAQ) system located on the surface of the Antarctic ice. 
The recorded data are digitised by a 14-bit \SI{240}{\mega\SPS} ADC on the mainboard to avoid data loss over the $\sim$\SI{2500}{\meter} powering cable. 
A more detailed description of the mainboard electronics can be found in~\cite{mainboard}.
In addition, three camera modules and an LED flasher system contribute to the calibration of the detector orientation and to ice property measurements which are aimed at reducing IceCube's systematic uncertainties~\cite{flasher}.

\section{Acceptance Testing} 

While the D-Eggs will be installed and operated as part of the IceCube Upgrade, prior to their transport to Antarctica, all D-Eggs undergo a practical examination of their hardware and software. A testing facility has been prepared at Chiba University, specifically to verify the functionality of the D-Eggs. The results presented here are from the first batch of D-Eggs to have finished the acceptance testing procedure.

\subsection{Experimental Setup and Motivation}

The tests performed balance two requirements --- hardware robustness and verifying physics capabilities. A core piece of this requirement is verification that the D-Eggs and their components are not damaged by large changes in temperature and can operate for sustained periods at cold temperatures. The regions of the South Pole ice in which the D-Eggs will be deployed experience temperatures ranging between around \SI{-10}{\celsius} and \SI{-40}{\celsius}, though during transport and storage they may experience temperatures much lower. To test the D-Eggs at typical operating temperatures, a large freezer capable of reaching an ambient temperature of \SI{-60}{\celsius} has been installed at the testing facility in Chiba. While being operated, the D-Egg mainboard and high voltage supplies produce heat such that the D-Eggs reach a resting temperature around \SI{-25}{\celsius} with the freezer set to \SI{-40}{\celsius}. Given the size of the freezer, 16 D-Eggs can be housed and tested simultaneously.

To verify the physics capabilities, the PMTs inside the D-Eggs must be tested. Testing the PMTs prior to integration into the D-Eggs has already been well-studied in~\cite{degg_performance} and the results shown here focus exclusively on signals from the PMTs that are digitised by the mainboard for fully integrated modules. In order to test the PMTs in an environment similar to that at the South Pole, optical fibres direct diffuse UV laser light into individual D-Egg-sized dark boxes inside the cold freezer.

The laser light (\SI{400}{\nm}) being directed into the boxes can be controlled in frequency, intensity, and emission pattern to examine properties particularly important for high energy neutrino events. This includes measuring the linearity of the PMT plus mainboard response to varying levels of light intensity and the ability to resolve two laser pulses in quick succession. An accurate understanding of the PMT plus mainboard linearity response is critical for correctly reconstructing the energy of high energy neutrino interactions, which produce large pulses of light. Additionally, the double-pulse waveform is a distinct, low background signature for tau neutrino charged current interactions in IceCube.

When the laser is not in operation, the dark environment can be used to measure properties of the PMT such as the gain or dark noise rate. Measurements of the D-Egg properties when not illuminated by light are critical for calibration purposes and understanding backgrounds observed during \textit{in-situ} operations.

\subsection{Testing Procedure}

16 D-Eggs were installed into individual light-tight boxes and tested at both room temperature and \SI{-40}{\celsius} (freezer temperature). Power is supplied to the D-Eggs from an external source located outside the freezer, and laser light is focused on an external optical bench before being distributed to each PMT inside the freezer via optical fibres. Upon exiting the fibre, the light is diffuse to a stop-size of several centimetres in diameter. The laser signal is controlled digitally via a function generator and the intensity is modulated by a 6-channel filter wheel. All data was collected using the D-Egg on-board electronics identical to those to-be-deployed at the South Pole. The PMTs are operated at a nominal gain of $10^7$, which produces single photon responses with a typical pulse height around $\SI{5}{\milli\volt}$. 

\section{Results}

\subsection{PMT Dark Noise Rate}
Dark noise describes backgrounds that did not originate from an external photon hitting the detector. This can result in the emission of an electron from the PMT cathode and be recorded by the mainboard. The sources of dark noise are thermionic cathode emission, PMT afterpulses, and radioactive processes within the glass components. The rate of thermionic emission is highly temperature dependent and can thus be reduced by measuring at freezing temperatures mimicking the temperatures in the deep glacial ice at the South Pole between \SI{-40}{\celsius} and \SI{-10}{\celsius}. Here the dark noise rate is measured at an ambient temperature of \SI{-40}{\celsius} and with a threshold of \num{0.25} times the average SPE peak amplitude. 

\begin{figure}[htb]
\center
\includegraphics[width=0.65\textwidth]{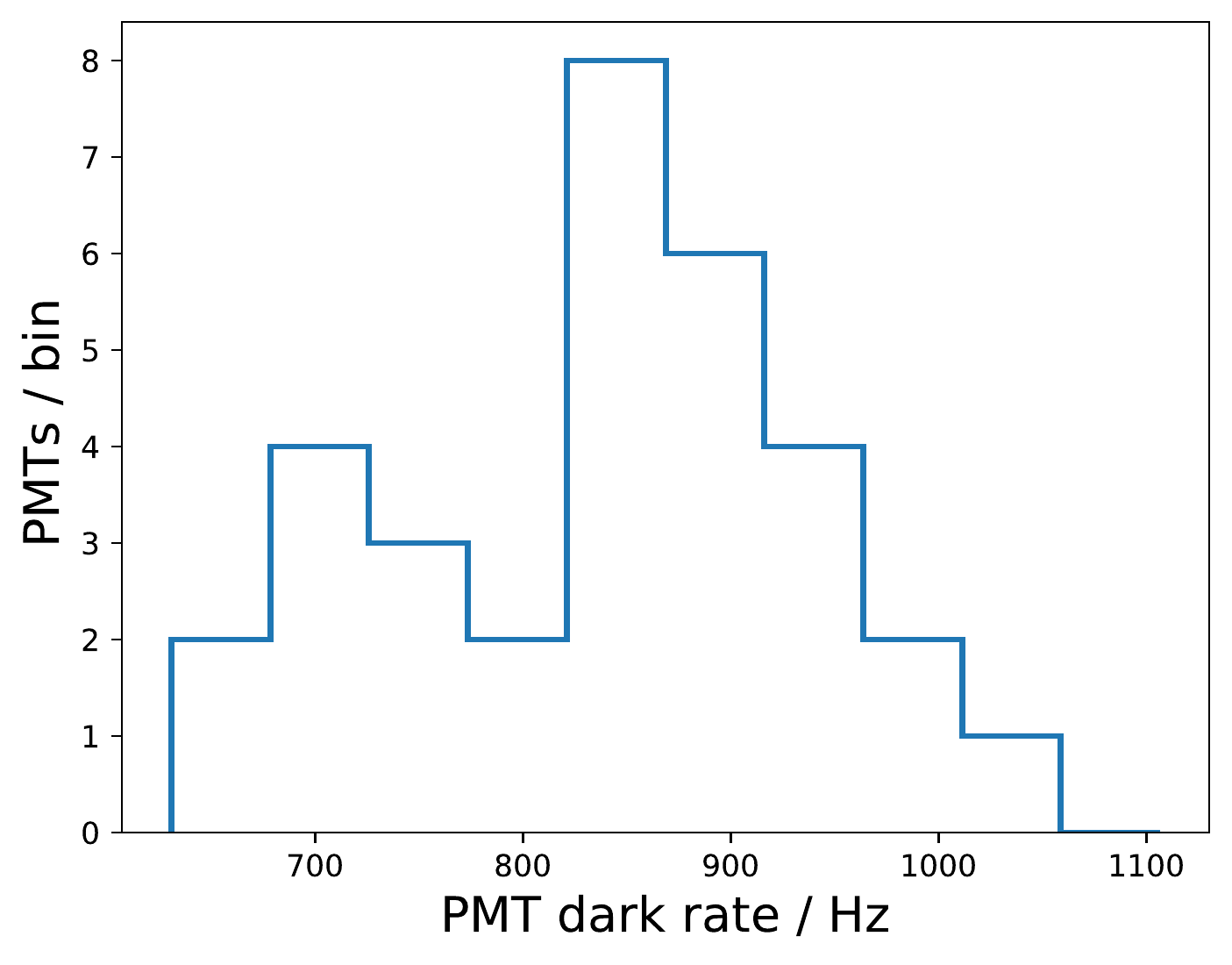}
\caption{Dark noise rate for \num{32} integrated D-Egg PMTs measured at an ambient temperature of \SI{-40}{\celsius}. Rates were measured with PMTs operating at $10^7$ gain, with a threshold of \num{0.25} times the average single photoelectron peak amplitude, and an artificial deadtime of \SI{100}{\nano\second} applied in software. The presented dark noise rates are corrected to the expected in-ice values.}
\label{fig:darkrate}
\end{figure}

In general, to obtain dark noise rates most comparable to operation at the South Pole, the PMT glass surface can be covered with black vinyl tape (described in more detail in~\cite{dom_paper}). By taping the glass surface, the observed dark noise rate decreases by roughly a factor of two. This is due to a difference in the refractive index between the glass--air and glass--ice boundaries. However, taping for all devices-under-tests is not feasible for acceptance testing due to logistics. Instead, the effect was measured using a subset of D-Eggs, and a calibration factor of \num{2.375} is applied to the dark noise rates presented here.
When operating the D-Eggs in a dark environment without an external light source a median dark noise rate per PMT of \SI{853}{\hertz} can be observed (see Fig.~\ref{fig:darkrate}). 

\FloatBarrier

\subsection{Single Photoelectron Charge and Waveform}

Characterizing the single photoelectron (SPE) waveform shape and charge distribution is especially important for event reconstructions. During detector operation, the PMT readout is triggered when a waveform amplitude exceeds \num{0.25} times the average SPE peak amplitude. This results in typical pulse heights around \SI{5}{\milli\volt}, where \SI{90}{\percent} of the charge is recorded within around \SI{15}{\nano\second} after the peak. 

\begin{figure}[htb]
\center
\includegraphics[width=.49\textwidth]{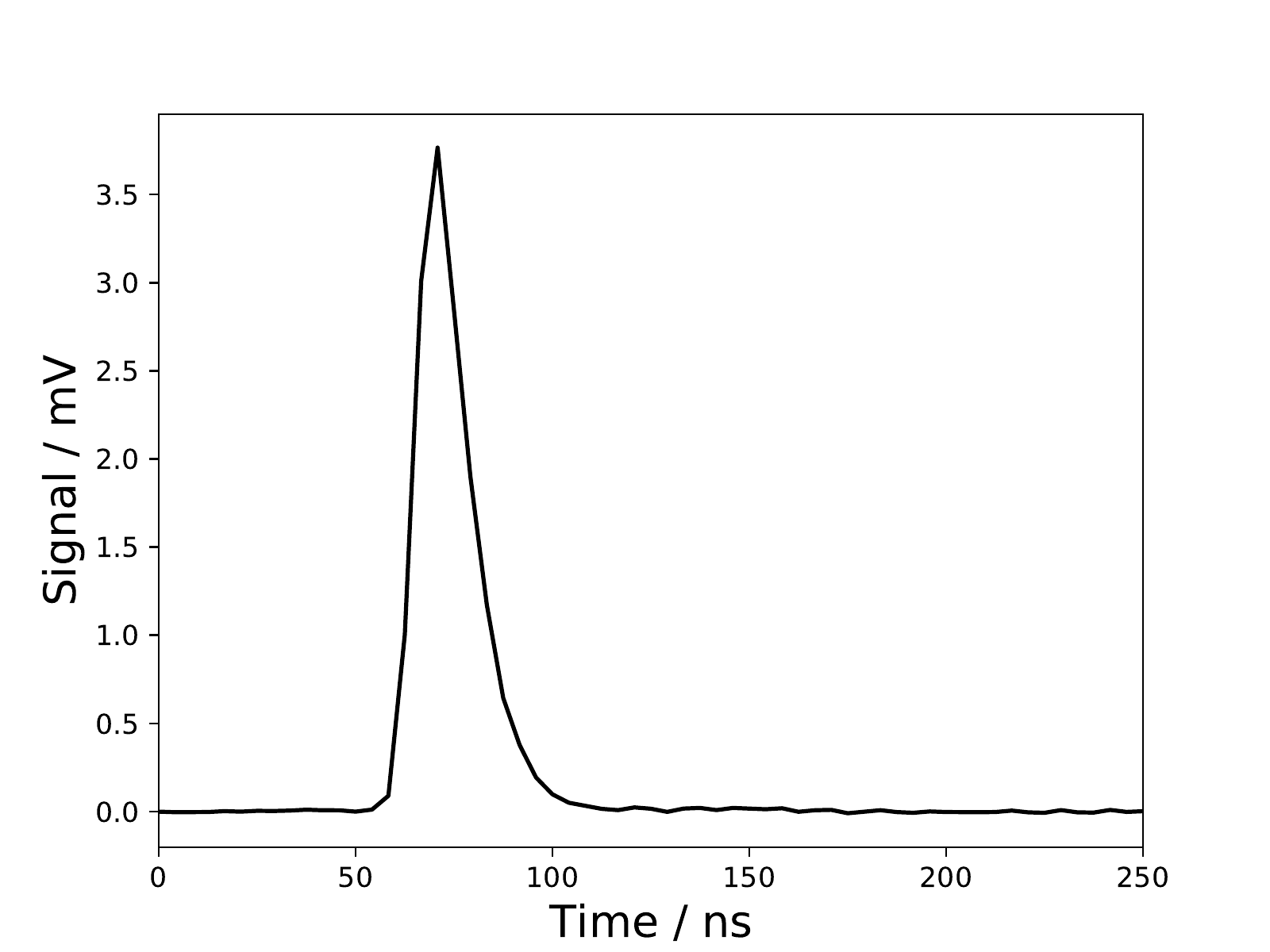}
\includegraphics[width=0.49\textwidth]{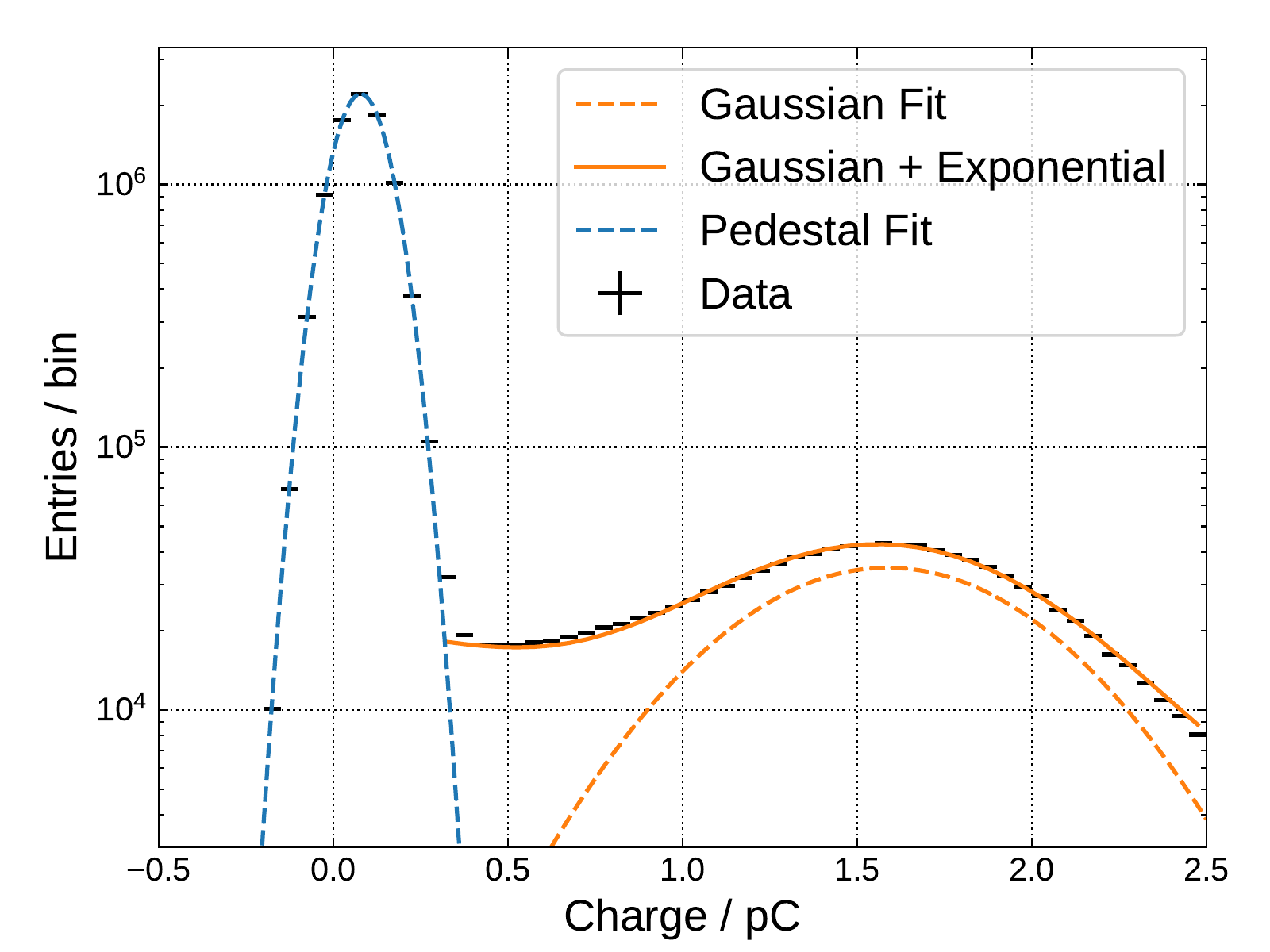}
\caption{Left: Average SPE waveform of $\sim$\num{2200} waveforms at $10^7$ gain. Right: Typical SPE charge distribution at $10^7$ gain with a fit by a model with a Gaussian plus an exponential term. The pedestal distribution has been subtracted before the fit.}
\label{fig:charge_dist}
\end{figure}

The charge distribution of those SPE waveforms is obtained by integrating the waveform within (\SI{-41.7}{\nano\second}, \SI{62.5}{\nano\second}) around the SPE peak. Here, waveforms are taken with a threshold of about \num{0.1} of the SPE peak height. Figure~\ref{fig:charge_dist} shows a typical SPE charge distribution for the D-Egg PMTs. The distribution is modelled by a Gaussian plus an exponential term to describe small charges close to zero after subtracting the pedestal distribution~\cite{dom_paper}. The charge resolution of the D-Egg PMT is about \SI{30}{\percent}, consistent with Gen1 performance.

\FloatBarrier

\subsection{D-Egg Linearity Response}

Measuring the linearity of the D-Egg involves recording the response of each PMT compared to a known amount of injected light. This is achieved using the 6-setting filter wheel, which reduces the laser light in a consistent and uniform way for all D-Egg PMTs. The filter wheel transmittance options are: \SI{5}{\percent},  \SI{10}{\percent},  \SI{25}{\percent},  \SI{34}{\percent},  \SI{50}{\percent}, and  \SI{100}{\percent}. The intensity of the laser was tuned such that  \SI{5}{\percent} transmittance corresponds to a signal observed by the PMT with a charge of a few 10s of photoelectrons and \SI{100}{\percent} results in single pulses of at-least \SI{200}{\PE}.

\begin{figure}[htb]
\center
\includegraphics[width=0.65\textwidth]{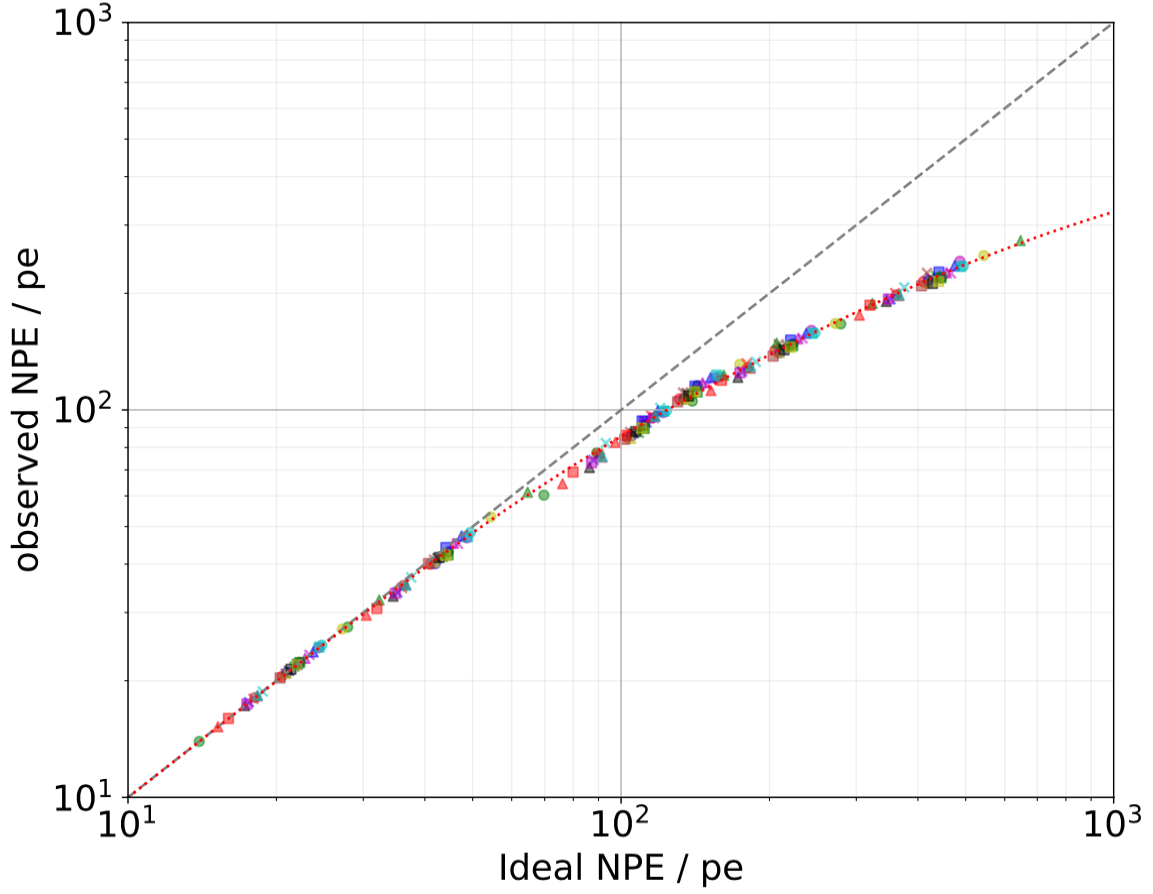}
\caption{Linearity response for \num{31} integrated D-Egg PMTs at \SI{-40}{\celsius}. Each PMT was measured at 6 filter transmittance settings, spanning the range of tens to hundreds of PE. The unique marker and colour combination correspond to a single PMT. The spread in ideal PE is driven mainly by variations in the optical fibre location and angle relative to the PMT. A 1:1 line (black) is used to guide the eye between the linear and non-linear region, and an arbitrary function (red) is fit to the data.}
\label{linearity_room_temp}
\end{figure}

Figure~\ref{linearity_room_temp} shows the linearity response measured at \SI{-40}{\celsius} for \num{31} of \num{32} PMTs -- one PMT was excluded due to damage to the optical fibre. The $y$-axis is the observed number of photoelectrons and the $x$-axis is the ideal number of photoelectrons which would be recorded by a perfectly linear system. The calculated ideal NPE are determined assuming that at a filter setting of \SI{5}{\percent} transmissivity, the observed and ideal NPE are equal, and that the ideal NPE scales linearly with the filter transmissivity: \SI{5}{\percent} to \SI{100}{\percent}. Given that the number of photoelectrons observed for the \SI{5}{\percent} filter is less than \SI{20}{\PE}, this should still be well-within the linear response region. At values around \SI{150}{\PE} the divergence from the linear response becomes obvious and all PMTs appear to be uniformly described by a single function even in the non-linear region. The departure of the system from the linear regime is driven by a combination of the PMT plus mainboard front-end signal shaper behaviour.

\FloatBarrier

\subsection{Double-Pulse Identification}

\begin{figure}[htb]
\center
\includegraphics[width=0.65\textwidth]{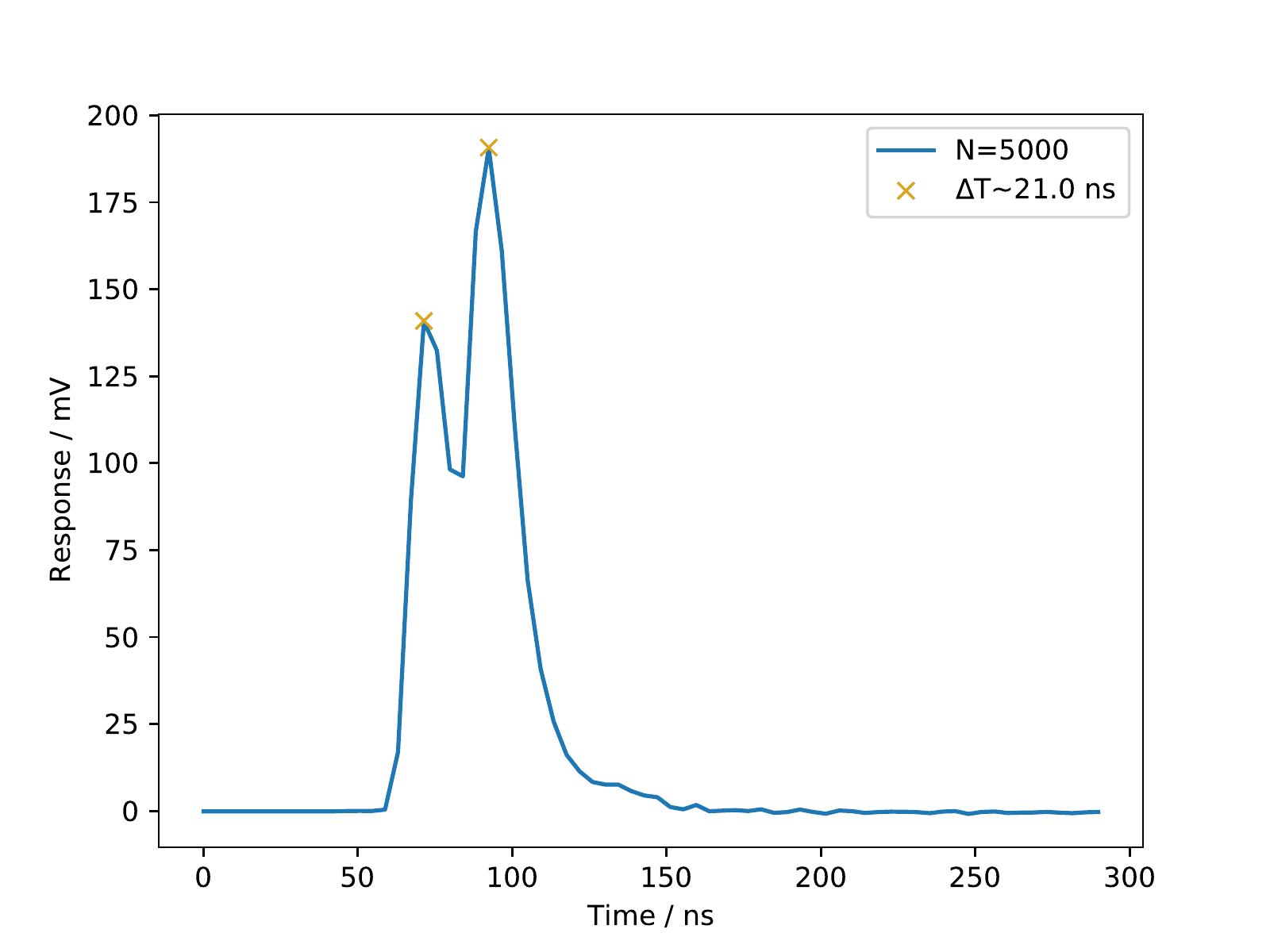}
\caption{Representative average of \num{5000} double-pulse waveforms for a D-Egg PMT. The double pulse structure is clearly visible and can be identified from a simple peak-finding algorithm (orange crosses). The time between the two pulses is consistent with the laser pulse interval.}
\label{double_pulse_cold_temp}
\end{figure}

Identification of two pulses separated by only a few nano-seconds is a strong indicator of high energy tau neutrino charged current interactions. To create a clean double-pulse signal, a function generator supplies the laser with a trigger for two pico-second width bursts separated by a \SI{20}{\nano\second} interval and then waits \SI{1}{\micro\second} between subsequent sets of pulses. This creates an observed waveform, the average of \num{5000} waveforms is seen in Fig.~\ref{double_pulse_cold_temp}, where the orange crosses indicate peaks located by a simple peak-finding algorithm.

For all PMTs the double-pulse structure could be extracted using the peak-finding algorithm and shows two clearly separated pulses (in time) identifiable by eye. The timing separation between the pulses is consistent to within half the size of the mainboard timing bins ($\sim$ \SI{4.2}{\nano\second}). Note, the \SI{20}{\nano\second} pulse separation is not the expected performance limit for the D-Eggs, but instead simple verification which aligns with hardware requirements. Thus, this verification setup conservatively probes the tau neutrino energy down to around \SI{40}{\tera\electronvolt}.

\FloatBarrier

\section{Conclusion}

While acceptance testing of D-Eggs is only just beginning, the D-Eggs sampled here all produced results which met or exceeded our initial testing requirements. This includes general operation of the electronics for a sustained period (one week or more) at \SI{-40}{\celsius}, as well as acceptable results for the PMT SPE, dark rate, linearity, and double-pulse responses. With over \num{300} more D-Eggs to be tested within the next year, the statistics available from these acceptance tests will inform IceCube on in-ice calibration procedures, in addition to building representative models for the D-Egg modules required for up-coming physics analyses.

\clearpage
\section*{Full Author List: IceCube Collaboration}


\scriptsize
\noindent
R. Abbasi$^{17}$,
M. Ackermann$^{59}$,
J. Adams$^{18}$,
J. A. Aguilar$^{12}$,
M. Ahlers$^{22}$,
M. Ahrens$^{50}$,
C. Alispach$^{28}$,
A. A. Alves Jr.$^{31}$,
N. M. Amin$^{42}$,
R. An$^{14}$
% ...
\\

\noindent
% ...
$^{12}$ Universit{\'e} Libre de Bruxelles, Science Faculty CP230, B-1050 Brussels, Belgium \\
% ...
$^{14}$ Department of Physics and Laboratory for Particle Physics and Cosmology, Harvard University, Cambridge, MA 02138, USA \\
% ...
$^{17}$ Department of Physics, Loyola University Chicago, Chicago, IL 60660, USA \\
$^{18}$ Dept. of Physics and Astronomy, University of Canterbury, Private Bag 4800, Christchurch, New Zealand \\
% ...
$^{22}$ Niels Bohr Institute, University of Copenhagen, DK-2100 Copenhagen, Denmark \\
% ...
$^{28}$ D{\'e}partement de physique nucl{\'e}aire et corpusculaire, Universit{\'e} de Gen{\`e}ve, CH-1211 Gen{\`e}ve, Switzerland \\
% ...
$^{31}$ Karlsruhe Institute of Technology, Institute for Astroparticle Physics, D-76021 Karlsruhe, Germany  \\
% ...
$^{42}$ Bartol Research Institute and Dept. of Physics and Astronomy, University of Delaware, Newark, DE 19716, USA \\
% ...
$^{50}$ Oskar Klein Centre and Dept. of Physics, Stockholm University, SE-10691 Stockholm, Sweden \\
% ...
$^{59}$ DESY, D-15738 Zeuthen, Germany \\
% ...

\subsection*{Acknowledgements}

\noindent
USA {\textendash} U.S. National Science Foundation-Office of Polar Programs,
U.S. National Science Foundation-Physics Division,
U.S. National Science Foundation-EPSCoR,
Wisconsin Alumni Research Foundation,
Center for High Throughput Computing (CHTC) at the University of Wisconsin{\textendash}Madison,
Open Science Grid (OSG),
Extreme Science and Engineering Discovery Environment (XSEDE),
Frontera computing project at the Texas Advanced Computing Center,
U.S. Department of Energy-National Energy Research Scientific Computing Center,
Particle astrophysics research computing center at the University of Maryland,
Institute for Cyber-Enabled Research at Michigan State University,
and Astroparticle physics computational facility at Marquette University;
Belgium {\textendash} Funds for Scientific Research (FRS-FNRS and FWO),
FWO Odysseus and Big Science programmes,
and Belgian Federal Science Policy Office (Belspo);
Germany {\textendash} Bundesministerium f{\"u}r Bildung und Forschung (BMBF),
Deutsche Forschungsgemeinschaft (DFG),
Helmholtz Alliance for Astroparticle Physics (HAP),
Initiative and Networking Fund of the Helmholtz Association,
Deutsches Elektronen Synchrotron (DESY),
and High Performance Computing cluster of the RWTH Aachen;
Sweden {\textendash} Swedish Research Council,
Swedish Polar Research Secretariat,
Swedish National Infrastructure for Computing (SNIC),
and Knut and Alice Wallenberg Foundation;
Australia {\textendash} Australian Research Council;
Canada {\textendash} Natural Sciences and Engineering Research Council of Canada,
Calcul Qu{\'e}bec, Compute Ontario, Canada Foundation for Innovation, WestGrid, and Compute Canada;
Denmark {\textendash} Villum Fonden and Carlsberg Foundation;
New Zealand {\textendash} Marsden Fund;
Japan {\textendash} Japan Society for Promotion of Science (JSPS)
and Institute for Global Prominent Research (IGPR) of Chiba University;
Korea {\textendash} National Research Foundation of Korea (NRF);
Switzerland {\textendash} Swiss National Science Foundation (SNSF);
United Kingdom {\textendash} Department of Physics, University of Oxford.

\end{document}